\documentclass[preprint,aps,tightenlines,showpacs,amsmath,amssymb,pra,superscriptaddress]{revtex4-1}
\usepackage{graphicx}
\usepackage{color}
\usepackage[nooneline,FIGTOPCAP,raggedright]{subfigure}
\usepackage{esint}
\usepackage{xparse}
\usepackage{soul}

\makeatletter
 
 \@ifundefined{textcolor}{}
 {%
   \definecolor{BLACK}{gray}{0}
   \definecolor{WHITE}{gray}{1}
   \definecolor{RED}{rgb}{1,0,0}
   \definecolor{GREEN}{rgb}{0,1,0}
   \definecolor{BLUE}{rgb}{0,0,1}
   \definecolor{CYAN}{cmyk}{1,0,0,0}
   \definecolor{MAGENTA}{cmyk}{0,1,0,0}
   \definecolor{YELLOW}{cmyk}{0,0,1,0}
 }

\NewDocumentCommand{\sotwo}{O{red}O{black}+m}
    {%
        \begingroup
        \setulcolor{#1}%
        \setul{-.5ex}{.4pt}%
        \def\SOUL@uleverysyllable{%
            \rlap{%
                \color{#2}\the\SOUL@syllable
                \SOUL@setkern\SOUL@charkern}%
            \SOUL@ulunderline{%
                \phantom{\the\SOUL@syllable}}%
        }%
        \ul{#3}%
        \endgroup
    }
\makeatother

\begin{document}

\title{Generation and stabilization of a three-qubit entangled W state in circuit QED via quantum feedback control}

\author{Shang-Yu Huang}
\author{Hsi-Sheng Goan}
\email{goan@phys.ntu.edu.tw}
\affiliation{Department of Physics and Center for Theoretical Sciences,
National Taiwan University, Taipei 10617, Taiwan}
\affiliation{Center for Quantum Science and Engineering, and National Center for Theoretical Sciences,
National Taiwan University, Taipei 10617, Taiwan}
\author{Xin-Qi Li}
\affiliation{Department of Physics, Beijing Normal University, Beijing 100875, China}
\author{G. J. Milburn}
\affiliation{Centre for Engineered Quantum Systems, School of
  Mathematics and Physics, The University of Queensland, St Lucia QLD 4072, Australia}

\date{\today}
\begin{abstract}
Circuit cavity quantum electrodynamics (QED) is proving to be a powerful
platform to implement quantum feedback control schemes due to the
ability to control superconducting qubits and microwaves in a
circuit. Here, we present a simple and promising quantum feedback
control scheme for deterministic generation and stabilization of a
three-qubit $W$ state in the superconducting circuit QED system. The
control scheme is based on continuous joint Zeno measurements of
multiple qubits in a dispersive regime, which enables us not only to
infer the state of the qubits for further information processing but
also to create and stabilize the target $W$ state through adaptive
quantum feedback control. We simulate the dynamics of the proposed
quantum feedback control scheme using the quantum trajectory approach
with an effective stochastic maser equation obtained by a polaron-type transformation method
and demonstrate that in the presence of moderate environmental
decoherence, the average state fidelity higher than $0.9$ can be
achieved and maintained for a considerable long time (much longer than
the single-qubit decoherence time). This control scheme is also shown
to be robust against measurement inefficiency and individual qubit
decay rate differences.
Finally, the comparison of the polaron-type transformation method to
the commonly used adiabatic elimination method to eliminate the cavity mode is presented.

\end{abstract}

\maketitle

\section{Introduction}
Entanglement is regarded as one of the key resources for various
applications in quantum information processing. 
While entanglement of bipartite systems is well understood \cite{Peres1996},
the characterization of multipartite entanglement is still an
interesting research topic. 
It has been shown \cite{Dur2000} that 
there are two
inequivalent non-biseparable classes of three-qubit entanglement states, the
Greenberger-Horne-Zeilinger (GHZ)class and the W
class, which cannot be transformed into each other by 
stochastic local operations and classical communications.
The W state is central as a resource in quantum
information processing and multi-party quantum communication
as its entanglement is persistent and robust even under
particle loss \cite{Dur2000,Guhne2008,Briegel2001}. 
The three-qubit entangled W state has been
experimentally generated and demonstrated in systems of trapped ions
\cite{Haffner2005}, optical photons \cite{Eibl2004},
superconducting phase qubits \cite{Neeley2010}, and coupled nonlinear oscillator arrays \cite{Gangat13}.

Circuit QED system
\cite{Blais2004,Wallraff2004,Wallraff2005,Schuster2005,Gambetta2006,Blais2007,Gambetta2007,Houck2008,Bonzom2008,Boissonneault2009,Filipp2009,DiCarlo2010,PhysRevA.81.062325,Leek2010,Filipp2011,DiCarlo09,Reed12,Fedorov12} in which superconducting qubits based on Josephson junctions
are coupled to a high-Q microwave transmission line resonator
acting as a quantum bus has been demonstrated to be a promising
solid-state quantum computing architecture.  
Due to the great controllability of the superconducting qubits and
microwaves in the circuit system, the circuit QED system, a solid-state analogy of quantum optics
cavity QED, also has excellent potential as a platform for quantum
control--especially quantum feedback control--experiments \cite{Nadav08,Vijay2011,Vijay2012,Murch12,Murch1301,Murch1305,Riste13}. 
In this paper, we present a simple and promising quantum feedback control
scheme for deterministic generation and stabilization of a
three-qubit $W$ state in the superconducting circuit QED system. 

Generation and manipulation of entangled
states are important tasks of quantum information processing.
Besides the scheme based on unitary dynamics to generate entangled states
\cite{Neeley2010,0957-4484-21-27-274015,Spilla2012}, there are proposals of 
entanglement generation by measurement \cite{PhysRevA.79.052328,Bishop2009,Lalumi`ere2010}.
Although measurements can generate entangled states that are
otherwise difficult to obtain, the specific or target entangled states 
created are primarily probabilistic.  
Furthermore, the measurement-alone approach cannot stabilize and
protect the generated entangled states from deterioration.

One possible way to resolve this problem is to employ the technique of
quantum feedback control \cite{Wiseman1993,Wiseman09,Carvalho2007,Sarovar05,Tornberg2010,Liu2010,Feng2011}. There have been proposals of using quantum
feedback control to stabilize
and generate two-qubit Bell states in circuit QED \cite{Sarovar05,Liu2010,Keane12}.  Rist\`{e} et al. \cite{Riste13}
recently presented an experimental demonstration of a superconducting two-qubit Bell state produced by feedback 
based on parity measurements. The case for three-qubit
entangled GHZ in circuit QED has also been investigated \cite{Feng2011}, but a
somewhat complicated method of an {\em alternate-flip-interrupted} Zeno
scheme and quantum feedback control technique with efficient
measurement and rapid single-qubit rotations are required
to produce and maintain the {\em pre-GHZ state} with high fidelity. 
However, how to generate and stabilize the other inequivalent
class of three-qubit states, namely the W state, in circuit QED has,
to our knowledge, not been reported. 

Here, we present a simple measurement and feedback control 
scheme that is feasible with current circuit QED technology to 
produce and stabilize the W state of  
$\left|W^{-}\right\rangle=(|100\rangle+|010\rangle+|001\rangle
)/\sqrt{3}$. 
Our scheme does not assume fast single-qubit
rotations and is robust against measurement inefficiency and
individual qubit decay rate differences.
A successful experimental implementation and realization of using
  quantum feedback control to generate and stabilize a multi-qubit entangled W
  state as presented here will be an impressive demonstration in
  circuit QED experiments. 
Moreover, previous investigations were performed in a
parameter regime of a strongly damped resonator cavity so that one can
adiabatically eliminate the 
cavity mode by enslaving the cavity to qubit dynamics \cite{Milburn1994,Brun2003,Wang2005,Sarovar05,Liu2010}.
Here, we go beyond this so-called bad-cavity limit by using a
polaron-type transformation \cite{Gambetta2008, Lalumi`ere2010} to trace out the cavity mode in our
analysis. This allows us to work in a parameter regime in which the
W state can be maintained with 
higher fidelity.   
The obtained effective (stochastic) master equation for the qubit degrees of
freedom provides us with more intuitive understanding and physical
insight into the qubit dynamics of the continuous quantum measurement and
quantum feedback control process. 
Note that our method can be extended straightforwardly to the
generation and stabilization of a $N$-qubit W-type state that is a
quantum superposition with equal expansion coefficients of all
possible pure states in which exactly one of the qubits is in an excited state $|1\rangle$, while all other ones are in the ground state $|0\rangle$.

The paper is organized as follows. We describe a three-qubit circuit
QED setup and its corresponding model Hamiltonian in Sec.~\ref{sec:model}.  The
procedure of using polaron-type transformation to eliminate cavity field to obtain an effective
master equation for the qubit degrees of freedom alone conditioned on
continuous homodyne detection is also presented in this section. The
quantum feedback control strategy to generate and stabilize the $W$
state of  
$\left|W^{-}\right\rangle=(|100\rangle+|010\rangle+|001\rangle
)/\sqrt{3}$ is described in Sec.~\ref{sec:QFC}. The results of
the average fidelity for the generation and stabilization of the
$\left|W^{-}\right\rangle$ state are presented in Sec.~\ref{sec:fidelity}. 
The dependence of the average fidelity on the qubits' decay rates $\gamma_j$,
dispersive coupling strength $\chi$, probe beam amplitude
$\epsilon$, feedback strength $f$, and measurement efficiency
$\eta$ are discussed. In Sec.~\ref{sec:comparison}, we
compare the polaron-type transformation method with the adiabatic
elimination method to eliminate the cavity mode. A short conclusion is
given in Sec.~\ref{sec:conclusion}.

\section{System: Hamiltonian and stochastic master equation}
\label{sec:model}

\begin{figure}
\centering
\includegraphics[angle=90,width=0.65\columnwidth]{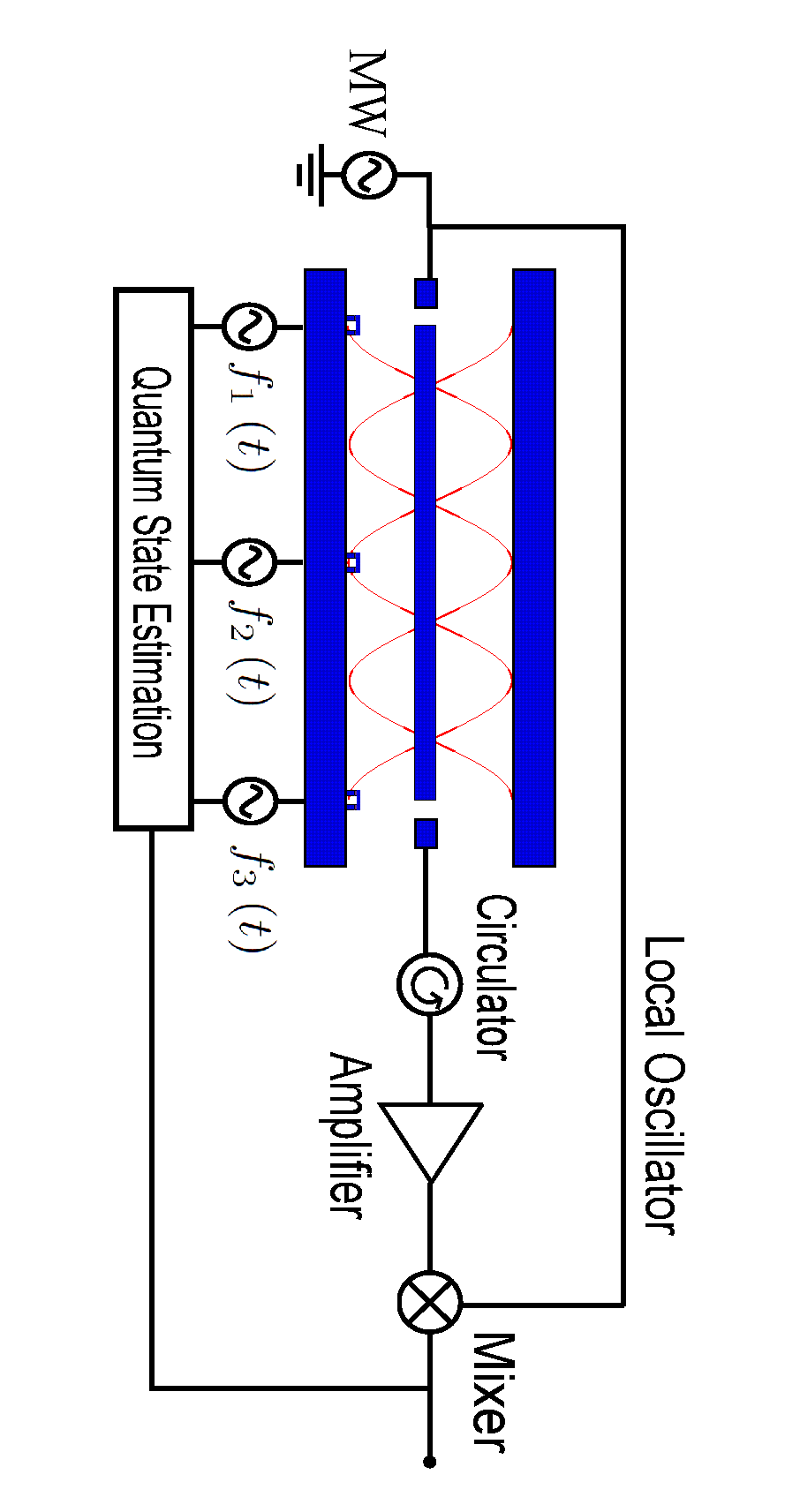}
\caption{\label{fig:setup} (Color online) Schematic illustration of 
  three qubits in circuit QED quantum feedback control setup (MW:
  microwave drive).}
\end{figure}

We consider a circuit QED setup in which three Cooper pair boxes
considered as qubits are coupled to a common field of a
one-dimensional microwave transmission line resonator (TLR) treated as
a cavity (see Fig.~\ref{fig:setup}).
The system can be described well by the Tavis-Cummings model
\cite{Blais2004,Fink2009,Liu2010,Feng2011,Lalumi`ere2010,Wu} and the Hamiltonian
driven by a measurement signal is described by 
\begin{eqnarray}
H & = & \omega_{r}a^{\dagger}a+\epsilon\left(a e^{i\omega_d t}+a^{\dagger}e^{-i\omega_d t}\right)\nonumber \\
 &  & +\sum_{j}\left[\frac{\Omega_{j}}{2}\sigma_{j}^{z}+g_{j}\left(\sigma_{j}^{-}a^{\dagger}+\sigma_{j}^{+}a\right)\right].\label{eq:Hamiltonian}
\end{eqnarray}
Here, the operators $\sigma_{j}^{-}$$\left(\sigma_{j}^{+}\right)$
and $a$$\left(a^{\dagger}\right)$ are respectively the lowering
(raising) operators 
of the $j$th qubit and the microwave inside the cavity, $\Omega_{j}$ 
is the transition frequency of the $j$th qubit, $\omega_{r}$ is the
cavity frequency, $g_{j}$ is the strength of
the $j$th qubit interacting with the cavity field, and $\epsilon$ and
$\omega_d$ are the amplitude and 
frequency of the measurement drive. 
In the dispersive
regime, where $\left|\Delta_{j}\right|=\left|\Omega_{j}-\omega_{r}\right|\gg g_{j}$,
we can eliminate the direct qubit-resonator coupling by
using the unitary transformation \cite{Blais2004} 
\begin{equation}
U=\exp\left[\sum_{j}\lambda_{j}\left(\sigma_{j}^{+}a-\sigma_{j}^{-}a^{\dagger}\right)\right].\label{XformOp}
\end{equation}
Keeping terms in the Hamiltonian up to second order in the small
parameter $\lambda_{j}=g_{j}/\Delta_{j}$ and moving to a frame
rotating with the measurement signal frequency $\omega_d$ for the cavity
field and qubits, we obtain
the Hamiltonian \cite{Lalumi`ere2010}
\begin{eqnarray}
H_{\text{eff}} & = &\left[\delta_{r}+\sum_{j}\chi_{j}\sigma_{j}^{z}\right]a^{\dagger}a+ \epsilon\left(a+a^{\dagger}\right)+\sum_{j}\epsilon\lambda_{j}\sigma_{j}^{x}\nonumber \\
 &  &
 +\sum_{j}\frac{\tilde{\Omega}_{j}}{2}\sigma_{j}^{z}+\sum_{j>i}J_{ij}^{q}\left(\sigma_{i}^{-}\sigma_{j}^{+}+\sigma_{i}^{+}\sigma_{j}^{-}\right).
\label{eq:effHamiltonian}
\end{eqnarray}
Here, the detuning frequency between cavity and the measurement drive 
$\delta_{r}=\omega_{r}-\omega_d$,
the dispersive coupling strength $\chi_{j}=g_{j}\lambda_{j}=g_{j}^2/\Delta_{j}$,
the dispersive-shifted qubit frequency $\tilde{\Omega}_{j}=\Omega_{j}-\omega_d+\chi_{j}$,
and the strength of qubit-qubit interaction mediated by the cavity
field
$J_{ij}^{q}=g_{i}g_{j}\left[\left(1/\Delta_{i}\right)+\left(1/\Delta_{j}\right)\right]/2$. 
One can see that in this dispersive limit, the qubit-resonator
interaction induces a qubit-state-dependent shift on the resonator
frequency. If we set $\delta_{r}=\omega_{r}-\omega_d=0$, i.e., the
driving frequency to be in resonance with the cavity frequency, the
measurement of the resonator frequency shift can be translated into the
measurement of the phase
shift between the incident and transmitted 
microwave drives.
Thus the information about the qubit state can be inferred
from the homodyne signal coming from the transmitted microwave
through the cavity or TLR. 
 The optimal signal to noise
  ratio for single-qubit dispersive readout is achieved for
$2\chi =\kappa$ \cite{Gambetta2008} while it is a little bit involved when 
multi-qubit joint measurement is considered [see discussions related
to Eqs.~(\ref{eq:jointmeasOp}) and (\ref{eq:JMOp2}) and to the red
dashed curve in Fig.~\ref{fig:fine}(a)].

The evolution equation 
for the density matrix of the qubit and
cavity system conditioned on the continuous homodyne detection in the joint rotating frame can be written as \cite{Wiseman1993,Wiseman09}
\begin{eqnarray}
\dot{\rho}_{c} & =& -\imath\left[H_{\text{eff}},\rho_{c}\right]+\sum_{j}\gamma_{j}\mathcal{D}\left[\sigma_{j}^{-}\right]\rho_{c}\nonumber \\
 &  & +\kappa\mathcal{D}\left[a \right]\rho_{c}
+\sqrt{\kappa\eta}\mathcal{H}[ae^{-i\phi}]\rho_{c}\xi(t),\label{eq:ME}
\end{eqnarray}
where the effect of the baths on the system describing the qubit and
cavity decays is denoted by the decoherence superoperator terms given in the
Lindblad form 
\begin{equation}
\mathcal{D}\left[c\right]\rho=c\rho c^{\dagger}-\frac{1}{2}\left(c^{\dagger}c\rho+\rho c^{\dagger}c\right),\label{eq:Lindblad}
\end{equation}
and $\kappa$ and $\gamma_i$ are respectively the cavity and individual
qubit decay rates. The last term in the conditional master equation
(\ref{eq:ME}) is the homodyne measurement unraveling
term that describes the back action and stochastic nature of the quantum
measurements, $0\le \eta\le 1$ is the measurement efficiency
($\eta=1$ corresponds to a perfect detector
or efficient measurement, and $\eta<1$ represents the fraction
of detections which are actually registered by the detectors), $\phi$ is the phase of the
local oscillator that is mixed with the transmitted microwave in the
homodyne measurement, and the measurement superoperator 
\begin{equation}
\mathcal{H}\left[c\right]\rho=c\rho+\rho c^{\dagger}-\left\langle c+c^{\dagger}\right\rangle \rho,\label{eq:MeasOp}
\end{equation}
where $\left\langle c\right\rangle =\text{tr}\left(\rho c\right)$
means the quantum average of the operator $c$.
The stochastic nature of the random measurement outcomes is
characterized by $\xi(t)$, a Gaussian white noise with the ensemble
average properties of $E[\xi(t)]=0$, $E[\xi(t)\xi(t')]=\delta(t-t')$,
where $E[\cdots]$ denotes an ensemble average over different
realizations of the noise. The use of a Gaussian white noise term here
assumes that the local oscillator has no more noise than a coherent
state, a good assumption at microwave frequencies. 
The measured homodyne current (in units of frequency) is proportional to 
\begin{equation}
I(t)=\kappa\eta\langle a\, e^{-i\phi}+a^{\dagger}\, e^{i\phi}\rangle+\sqrt{k\eta}\xi(t).
\label{eq:homodyne_I}
\end{equation}
Although Eq.~(\ref{eq:ME}) can be used to study the conditional
dynamics and measurement backaction, it provides little direct insight about
how the evolution of the qubits depends on the continuous measurement
outcomes. However, if the cavity field can be traced out and an 
effective (stochastic) master equation for the qubit degrees of
freedom only can be obtained, more intuition and understanding to the
qubit dynamics 
of the continuous quantum measurement process can be
gained and thus help
facilitate the successful development and design of
further manipulation and control strategies for the qubit system, e.g., the 
quantum feedback control strategy presented later in this article.

To obtain the effective stochastic master equation for the qubits'
degrees of freedom only, 
a common method is the so-called adiabatic elimination procedure valid 
in the limit where the damping of the cavity is much larger than
both the dispersive coupling strengths and the qubits' decay rates, i.e.,
$\kappa\gg (\chi_i$ and $\gamma_i)$. 
Here, we go beyond this limit and use a polaron-type transformation
\cite{Gambetta2008, Lalumi`ere2010} 
to trace out the cavity field. 
We will compare these two approaches in Sec.~\ref{sec:comparison}.
The polaron-type approach assumes only that the state of the qubits
varies slowly within the measurement time during which
the cavity field evolves to a 
steady coherent state depending on the
qubit state. 
This assumption can be justified if the cavity field decay rate is
much faster than 
the qubit decay rate $\kappa\gg2\gamma_{j}$.  
In this case, the unconditional master equation, i.e.,
Eq.~(\ref{eq:ME}) averaged over the white noise process,  indicates that a coherent state remains a coherent state with the
amplitude $\alpha_x$ of the cavity coherent state $|\alpha_x\rangle$
at $\delta_{r}=\omega_{r}-\omega_d=0$
satisfying
\begin{equation}
\dot{\alpha}_{x}=-\imath\chi_{x}\alpha_{x}-\imath\epsilon-\frac{\kappa}{2}\alpha_{x}
\label{eq:Coherent}
\end{equation}
when the qubits are in a basis state  $\left| x \right\rangle
= \left| ijk \right\rangle$, where $i,j,k\in\left\{ 0,1\right\} $,  $\left|0\right\rangle $ and
$\left|1\right\rangle $ represent respectively the ground and excited states of a single qubit, and $\chi_{x}=\left\langle
  x\right|\sum_{j}\chi_{j}\sigma_{j}^{z}\left|x\right\rangle$. 
Then the elimination of the cavity (TLR) degrees of freedom is carried out by 
going to a frame 
defined by the transformation \cite{Gambetta2008, Lalumi`ere2010}
\begin{equation}
\mathbf{P}\left(t\right)=\sum_{x}\Pi_{x}D\left[\alpha_{x}\left(t\right)\right]\label{eq:Polaron}
\end{equation}
with $D\left[\alpha\right]$ the displacement operator of the TLR,
\begin{equation}
D\left[\alpha\right]=\exp\left[\alpha a^{\dagger}-\alpha^{*}a\right],\label{eq:Polaron01}
\end{equation}
and $\Pi_{x}=\left| x \right\rangle\left\langle x\right|$ are projection
operators onto the respective basis (logical) states of the three-qubit
Hilbert space.
In this transformed reference frame, 
the cavity field is displaced to start from a vacuum state with zero photon,
i.e., $D\left[\alpha_x\right]|0\rangle_{TLR}=|\alpha_x\rangle_{TLR}$.
For simplicity, we take $\Omega_{1}=\Omega_{2}=\Omega_{3}=\Omega$
and $g_{1}=g_{2}=g_{3}=g$. This implies that we assume three identical
qubits with one wavelength separation apart in the TLR cavity (see Fig.~\ref{fig:setup}).
This assumption also helps generate the $|W^{-}\rangle$ state
by continuous measurements
as the $|W^{-}\rangle$ state in this case is a simultaneous eigenstate of the
system Hamiltonian (i.e., when without consideration of the qubits' decay) 
and the homodyne measurement operator [see Eq.~(\ref{eq:JMOp2}) and further discussion below it].
Then following the calculations in
Refs.~\onlinecite{Gambetta2008,Lalumi`ere2010}, we obtain an effective
master equation for the qubits' degrees of freedom alone conditioned on
continuous homodyne detection as
\cite{Gambetta2008,Tornberg2010,Lalumi`ere2010}
\begin{eqnarray}
d\rho_{c}\left(t\right)/dt & = & \mathcal{L}\rho_{c}\left(t\right)+\sqrt{\kappa\eta}\mathcal{H}\left[c_{\phi}\right]\rho_{c}\left(t\right)\xi\left(t\right)-\imath\sqrt{\kappa\eta}\left[c_{\phi-\pi/2},\rho_{c}\left(t\right)\right]\xi\left(t\right),\label{eq:SME}
\end{eqnarray}
where $\mathcal{L}\rho_{c}$ is given by 
\begin{eqnarray}
\hspace{-0.4cm}\mathcal{L}\rho_{c}& = &
-\imath\left[\sum_{j}\frac{\chi}{2}\sigma_{j}^{z}
+\sum_{j}\epsilon\lambda(\sigma_{j}^{+}e^{i\Delta t}+\sigma_{j}^{-}e^{-i\Delta t})+\sum_{j>i}\chi\left(\sigma_{i}^{-}\sigma_{j}^{+}+\sigma_{i}^{+}\sigma_{j}^{-}\right),\rho_c\right]
\nonumber \\
 &  &
 +\sum_{j}\gamma_{j}\mathcal{D}\left[\sigma_{j}^{-}\right]\rho_c
+\kappa\mathcal{D}\left[\sum_{j}\lambda\sigma_{j}^{-}\right]\rho_c+\sum_{xy}\left(\Gamma_{d}^{xy}-\imath
   A_{c}^{xy}\right)\Pi_{x}\rho_c\Pi_{y}.
\label{eq:MS}
\end{eqnarray}
In writing Eq.(\ref{eq:MS}), we have transformed to a frame rotating
with the qubits' transition frequency $\Omega$. This is also a
suitable frame for applying an additional microwave drive with a
frequency in resonance with the qubits' transition frequency in order to
coherently control the qubits as discussed in the context of
quantum feedback control in the next section.  
As a result, the $\sigma_x$ term
in Eq.~(\ref{eq:effHamiltonian}) now acquires time-dependent
oscillating factors
with frequency $\Omega-\omega_d=\Delta$ (as we have set
$\omega_d=\omega_r$) in the commutator of the first term in Eq.~(\ref{eq:MS}).  
The third term in Eq.~(\ref{eq:MS}) represents the Purcell effect at the damping rate $\kappa\lambda^{2}$
which can be reduced by operating the qubits in the dispersive
regime \cite{Boissonneault2009,Houck2008,Filipp2011}, while the fourth 
term contains both the measurement-induced dephasing $\left(\Gamma_{d}^{xy}\right)$
and the ac Stark shift $\left(A_{c}^{xy}\right)$ given by 
\begin{eqnarray}
\Gamma_{d}^{xy} & = & \left(\chi_{x}-\chi_{y}\right)\text{Im}\left[\alpha_{x}\alpha_{y}^{*}\right],\label{eq:dephasing}\\
A_{c}^{xy} & = & \left(\chi_{x}-\chi_{y}\right)\text{Re}\left[\alpha_{x}\alpha_{y}^{*}\right].\label{eq:acStark}
\end{eqnarray}
The last two terms of Eq.~(\ref{eq:SME}) come from the last term of the
conditional master equation (\ref{eq:ME}) in the displaced
polaron-type frame, in which the cavity field is transformed into 
\begin{equation}
a\, e^{-i\phi}\to \sum_x \Pi_x\alpha_x e^{-i\phi}=c_{\phi}-ic_{\phi-\pi/2}.
\label{eq:Ha}
\end{equation}
The measured homodyne current from  Eq.~(\ref{eq:homodyne_I}) becomes 
\begin{equation}
I_c\left(t\right)=\kappa\eta\left\langle c_{\phi}+c_{\phi}^{\dagger}\right\rangle _{c}\left(t\right)+\sqrt{\kappa\eta}\xi\left(t\right).\label{eq:current}
\end{equation}
Here the joint measurement operator $c_{\phi}$ is given by \cite{Gambetta2008,Tornberg2010,Lalumi`ere2010}
\begin{equation}
c_{\phi}=\frac{1}{2}\sum_{i,j,k=0}^{1}\sqrt{\Gamma_{ijk}\left(\phi\right)}\left(\sigma^{z}_{1}\right)^{i}\left(\sigma^{z}_{2}\right)^{j}\left(\sigma^{z}_{3}\right)^{k},\label{eq:jointmeasOp}
\end{equation}
 where 
\begin{equation}
\Gamma_{ijk}\left(\phi\right)=\left|\beta_{ijk}\right|^{2}\cos^{2}\left(\phi-\theta_{\beta_{ijk}}\right),\label{eq:MeasureRate}
\end{equation}
\begin{equation}
\beta_{ijk}=\frac{1}{4}\sum_{l,m,n=0}^{1}\left(-1\right)^{\vec{a}\cdot\vec{b}}\alpha_{lmn},\label{eq:amplitude}
\end{equation}
 where the vectors are $\vec{a}=\left(i,j,k\right)$ and $\vec{b}=\left(1-l,1-m,1-n\right)$, $\theta_{\beta}=\arg\left(\beta\right)$,
 and $\kappa\eta\Gamma_{ijk}\left(\phi\right)$
is the measurement rate for the polarization of $\left(\sigma^{z}_{1}\right)^{i}\left(\sigma^{z}_{2}\right)^{j}\left(\sigma^{z}_{3}\right)^{k}$. 
Note that the conditional stochastic master equation (\ref{eq:SME}) 
 after being averaged over all possible measurement
records reduces to the unconditional, deterministic
master equation, i.e., Eq.~(\ref{eq:SME}) but without its last two
unraveling noise terms.

The outcomes of the homodyne current (\ref{eq:current}) depend on the choice of the local oscillator
phase $\phi$. We would like to generate the entanglement state
$|W^{-}\rangle$ by quantum measurement. Thus we choose the phase to be $\phi=0$ such that
$|W^{-}\rangle$ state is one of the eigenstates of the measurement
operator \cite{Bishop2009,PhysRevA.81.062325,Spilla2012,Feng2011}
\begin{eqnarray}
c_{0}&=&\frac{3\sqrt{\Gamma_{0}}-\sqrt{\Gamma_{1}}}{2}\left(\Pi_{111}-\Pi_{000}\right)\nonumber\\
&&+\frac{\sqrt{\Gamma_{0}}+\sqrt{\Gamma_{1}}}{2}\left(\Pi_{011}+\Pi_{101}+\Pi_{110}-\Pi_{100}-\Pi_{010}-\Pi_{001}\right)\\
&=&\frac{\sqrt{\Gamma_{0}}}{2}\left(\sigma_{1}^{z}+\sigma_{2}^{z}+\sigma_{3}^{z}\right)-\frac{\sqrt{\Gamma_{1}}}{2}\sigma_{1}^{z}\sigma_{2}^{z}\sigma_{3}^{z},\label{eq:JMOp2}
\end{eqnarray}
where 
\begin{eqnarray}
\sqrt{\Gamma_0}&=&\sqrt{\Gamma_{100}}=\sqrt{\Gamma_{010}}=\sqrt{\Gamma_{001}}\nonumber\\
&=&\frac{1}{4}(\alpha_{111}-\alpha_{000}+\alpha_{110}-\alpha_{001}),
\label{gamma0}
\end{eqnarray}
and 
\begin{eqnarray}
\sqrt{\Gamma_1}&=&\sqrt{\Gamma_{111}}\nonumber\\
&=&\frac{1}{4}(\alpha_{000}-\alpha_{111}+3\alpha_{001}-3\alpha_{110}).
\label{gamma1}
\end{eqnarray}
In this case, there are four measurement 
outcomes of $\langle c_{0}+c_{0}^{\dagger}\rangle$:  
$3\sqrt{\Gamma_{0}}-\sqrt{\Gamma_{1}}$,
$\sqrt{\Gamma_{0}}+\sqrt{\Gamma_{1}}$, $-\sqrt{\Gamma_{0}}-\sqrt{\Gamma_{1}}$,
and $-3\sqrt{\Gamma_{0}}+\sqrt{\Gamma_{1}}$ which correspond respectively
to the all-qubit excited state $\left|111\right\rangle $, the
two-qubit excited states $\left\{\left|110\right\rangle,
  \left|101\right\rangle, \left|011\right\rangle \right\}$, the
single-qubit excited states $\left\{\left|001\right\rangle,
  \left|010\right\rangle, \left|100\right\rangle \right\}$, and the
ground state $\left|000\right\rangle$.
One can see from Eq.~(\ref{eq:Ha}) and the last two terms of
Eq.~(\ref{eq:SME}) that in addition to the measurement operator
$c_{0}$ providing the qubit state information, performing the homodyne measurement at
$\phi=0$ also produces a stochastic phase represented by the term
associated with 
\begin{eqnarray}
c_{-\pi/2}&=&\frac{3\sqrt{\Gamma_{2}}}{2}\left(\Pi_{111}+\Pi_{000}\right)\nonumber\\
&&-\frac{\sqrt{\Gamma_{2}}}{2}\left(\Pi_{011}+\Pi_{101}+\Pi_{110}+\Pi_{100}+\Pi_{010}+\Pi_{001}\right)\\
&=&\frac{\sqrt{\Gamma_{2}}}{2}\left(\sigma_{1}^{z}\sigma_{2}^{z}+\sigma_{2}^{z}\sigma_{3}^{z}+\sigma_{1}^{z}\sigma_{3}^{z}\right),
\end{eqnarray}
where 
\begin{eqnarray}
\sqrt{\Gamma_2}&=&\sqrt{\Gamma_{011}}=\sqrt{\Gamma_{101}}=\sqrt{\Gamma_{110}}\nonumber\\
&=&\frac{i}{4}(\alpha_{111}+\alpha_{000}-\alpha_{110}-\alpha_{001}).
\label{gamma2}
\end{eqnarray}
Note that $c_{-\pi/2}$ generates different relative phase
kicks only between two groups of states, i.e., between the group of 
$\left\{\left|111\right\rangle, \left|000\right\rangle \right\}$ and
the group of 
$\left\{\left|110\right\rangle,
  \left|101\right\rangle, \left|011\right\rangle, \left|001\right\rangle,
  \left|010\right\rangle, \left|100\right\rangle \right\}$. In other
words, no relative random phase kick between the
constituent basis states of the $|W^{-}\rangle$ state (thus no
additional unwanted dephasing between them) also helps
generate and stabilize the target $|W^{-}\rangle$ state.

When the coherent state amplitudes
are steady, the rates $\kappa\Gamma_{0}$,  $\kappa\Gamma_{1}$ and $\kappa\Gamma_{2}$ become
\begin{eqnarray}
\sqrt{\kappa\Gamma_{0}} & = & \sqrt{\Gamma_{\rm m}}\left[\frac{1+12\left(\chi/\kappa\right)^{2}}{1+40\left(\chi/\kappa\right)^{2}+144\left(\chi/\kappa\right)^{4}}\right], \label{eq:rate0}\\
 \sqrt{\kappa\Gamma_{1}}& = & \sqrt{\Gamma_{\rm m}}\left[\frac{24\left(\chi/\kappa\right)^{2}}{1+40\left(\chi/\kappa\right)^{2}+144\left(\chi/\kappa\right)^{4}}\right],\label{eq:rate1}\\
 \sqrt{\kappa\Gamma_{2}}& = & \sqrt{\Gamma_{\rm m}}\left[\frac{-4\left(\chi/\kappa\right)}{1+40\left(\chi/\kappa\right)^{2}+144\left(\chi/\kappa\right)^{4}}\right],\label{eq:rate2}
\end{eqnarray}
where $\Gamma_{\rm m}=64\epsilon^{2}\chi^{2}/\kappa^{3}$ is the
effective measurement rate obtained from the adiabatic elimination
method \cite{Sarovar05,Liu2010}.
In general, the measurement rate is related to the decoherence rate
as the decrease in the
off-diagonal element is associated with the gradual projection
onto one of the corresponding measurement eigenstates.
For example, in the steady state , the measurement rate 
$\Gamma_{\rm m}$ for efficiency $\eta=1$ is equal to twice of the decoherence
rate, i.e., $\Gamma_{\rm m}=2\Gamma_{\rm e}$, for the case by the
adiabatic elimination method [see Eq.(\ref{eq:aSME})] \cite{Liu2010}. 
Similarly, there is a relationship between the measurement rates
defined through 
Eq.~(\ref{eq:JMOp2}) and the corresponding decoherence rates in
Eq.~(\ref{eq:dephasing}). 
We may define for efficiency $\eta=1$ the ratio between them as 
\begin{equation}
R_{x,y}\equiv\frac{\kappa\left|\alpha_{x}-\alpha_{y}\right|^{2}}{\left(\chi_{x}-\chi_{y}\right)\text{Im}\left[\alpha_{x}\alpha^{*}_{y}\right]},
\end{equation}
where the numerator, $\kappa \left|\alpha_{x}-\alpha_{y}\right|^2$, is
the measurement rate to distinguish between two eigenstates
$|x\rangle$ and $|y\rangle$ of the measurement operator $c_0$
(also proportional to the separation between two measurement
outcomes corresponding to the states $|x\rangle$ and $|y\rangle$), and the denominator is the measurement-induced dephasing in Eq.~(\ref{eq:dephasing}).
When the coherent amplitudes have reached steady state, it can be shown
that
\begin{equation}
R_{000,111}=R_{100,011}=R_{010,101}=R_{001,110}=2.
\end{equation}
In other words,
 the measurement rates to distinguish between the states
 $\left|111\right\rangle\leftrightarrow\left|000\right\rangle$,
 $\left|011\right\rangle\leftrightarrow\left|100\right\rangle$,
 $\left|101\right\rangle\leftrightarrow\left|010\right\rangle$, and
 $\left|110\right\rangle\leftrightarrow\left|001\right\rangle$ are
 twice of their corresponding  measurement-induced dephasing rates.

We wish to apply the measurement-guided quantum feedback control to
generate and stabilize the $|W^{-}\rangle$ state, thus distinct values
of measurement current, which reveal quickly the information of
corresponding qubit states are favorable. 
The separations between
adjacent measurement outcomes are $2\sqrt{\Gamma_{0}}\pm
2\sqrt{\Gamma_{1}}$ that depend on the value of $\chi/\kappa$. 
Generally, larger separation between measurement outcomes implies
  quicker corresponding measurement eigenstate readout.
Let us focus on the smaller separation of
$2\sqrt{\Gamma_{0}}-2\sqrt{\Gamma_{1}}$ between the measurement
outcome of the
ground state and that of the $|W^{-}\rangle$ state (or the
single-qubit excited states). 
When the ratio $\chi/\kappa$ is decreased from $0$ to $-1$ , the
separation between the outcomes of the ground state and the
single-qubit excited state [see the red-dashed curve in
Fig.~\ref{fig:fine}(a)] increases initially, and then reaches a local
maximum around $\chi/\kappa=-0.11$. It then vanishes around
$\chi/\kappa=-0.29$ and later reaches another local maximum around 
$\chi/\kappa=-0.77$. However, a larger $\chi$ value leads to 
a larger damping rate,
$\kappa\lambda^{2}=\kappa\chi^2/g^2$,
of the Purcell effect which deteriorates the average 
fidelity to generate and stabilize the $|W^-\rangle$ state. As a
result, the fidelity at $\chi=-0.77\kappa$ is expected to be 
smaller than that at 
$\chi=-0.11\kappa$. We will show later that the average 
fidelity at $\chi=-0.77\kappa$ is also smaller than that at, say,
$\chi=-0.5\kappa$ even though its measurement outcome separation is
larger [see Fig.~\ref{fig:fine}\subref{fig:fine_chi} and the
discussion related to it in Sec.~\ref{sec:fidelity}]. 
Thus for the simulations presented in this article,
the dispersive coupling strength is chosen to be $\chi=-0.11\kappa$   
and/or $\chi=-0.50\kappa$. 
These are readily accessible parameter values
  \cite{Gambetta2006,Schuster07,Fink08,Bishop09}.

\begin{figure}
\centering
\includegraphics[width=0.45\columnwidth]{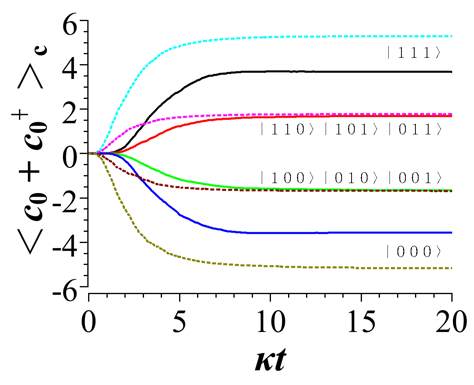}

\caption{\label{fig:1} (Color online) Measured homodyne currents 
$\langle c_{0}+c^\dagger_{0}\rangle_c$ ($1000$ realizations that
yield roughly the same steady outcome values are grouped and averaged) for an initial cavity state in a vacuum state and an initial
qubits' state in the separable state
$\left|\psi_{i}\right\rangle$ of Eq.~(\ref{eq:FactorState}) with the qubits'
decay rates $\gamma_j=0$. The four measurement outcomes
in solid lines are for the case of polaron-type transformation and they
correspond to the qubits collapsing respectively onto states
$\left|111\right\rangle$, $\left|W^{+}\right\rangle$,
$\left|W^{-}\right\rangle$, and $\left|000\right\rangle$ from the top
to the bottom, while the measurement outcomes
in dashed lines corresponds to the case of adiabatic elimination.
The parameters used are $\epsilon=2\kappa$, $\chi=-0.11\kappa$,
$g=10\kappa$, $\eta=1$, and $\gamma_j=0$.}
\end{figure}

Suppose we start to evolve the conditional master equation
(\ref{eq:SME}) with an 
initial state of the cavity in a vacuum state and the qubits in a
separable state 
\begin{equation}
\left|\psi_{i}\right\rangle =\frac{1}{\sqrt{2}}\left(\left|0\right\rangle +\left|1\right\rangle \right)_{1}\otimes\frac{1}{\sqrt{2}}\left(\left|0\right\rangle +\left|1\right\rangle \right)_{2}\otimes\frac{1}{\sqrt{2}}\left(\left|0\right\rangle +\left|1\right\rangle \right)_{3}.\label{eq:FactorState}
\end{equation}
Ideally, it is expected that the qubits under continuous measurements will collapse gradually onto one of the
eigenstates of the joint-qubit measurement operator $c_{0}$
stochastically in each individual realization. 
Indeed, by ignoring the decay rates
of the qubits, i.e., by setting $\gamma_{j}=0$,
the initial qubit state $\left|\psi_{i}\right\rangle$ of
Eq.~(\ref{eq:FactorState}) will collapse onto the states
$\left|111\right\rangle $ and $\left|000\right\rangle $ with
probability $0.125$ each, and onto the states
$\left|W^{+}\right\rangle=(|110\rangle+|101\rangle+|011\rangle
)/\sqrt{3}$ and
$\left|W^{-}\right\rangle=(|100\rangle+|010\rangle+|001\rangle
)/\sqrt{3}$ with probability $0.375$ each. 
This is shown in Fig.~\ref{fig:1} where the averaged measured currents 
$\langle c_{0}+c^\dagger_{0}\rangle_c$ 
obtained by categorizing and averaging $1000$ realizations that
yield roughly the same steady outcome values [as implied by Eq. (\ref{eq:FactorState})]
are presented.
One can see that after the time of about $5/\kappa$,  the
cavity field has evolved from the initial vacuum state to 
correspondingly distinguishable coherent states and four
distinct measurement outcomes (solid lines in Fig. \ref{fig:1}) are
observed. The measurement outcomes maintaining at certain values for a
considerably long time indicate that the qubits have collapsed onto
and stayed in the corresponding states of $\left|111\right\rangle$,
$\left|W^{+}\right\rangle$, $\left|W^{-}\right\rangle$ and
$\left|000\right\rangle$. 
However, this scheme of producing entangled $|W^{-}\rangle$ or
$\left|W^{+}\right\rangle$ states by measurement only
is probabilistic, and in the
presence of qubit relaxation, the probabilistically generated
entangled state will jump into other states.

\section{Entanglement creation and stabilization by quantum feedback control}

\subsection{Quantum feedback control strategy}
\label{sec:QFC}

To generate the $|W^{-}\rangle$ state deterministically and stabilize it
against the influence of the environments, we employ the adaptive quantum
feedback control technique based on quantum state estimation.
The conditional stochastic master equation with this kind of the feedback
control scheme becomes \cite{Sarovar05,Liu2010,Ahn2002}
\begin{eqnarray}
\dot{\rho}_{c}\left(t\right) & = &
\mathcal{L}\rho_{c}\left(t\right)+\sqrt{\kappa}\mathcal{H}\left[c_{0}\right]\rho_{c}\left(t\right)\xi\left(t\right)-\imath\sqrt{\kappa}\left[c_{-\pi/2},\rho_{c}\left(t\right)\right]\xi\left(t\right)
\nonumber \\
&&-i[H_{\rm fb}(t),\rho_c(t)].\label{eq:feedbackSME}
\end{eqnarray}
Here $H_{\rm fb}\left(t\right)$ is the feedback control Hamiltonian
 with control parameters designed from an estimation of
 $\rho_{c}\left(t\right)$.  The advantage of the quantum state
 estimation scheme is that the feedback control can be designed from
 an optimal control method to ensure the qubit system passing through
 the more efficient trajectory by optimizing the targeted objective or
 minimizing the cost function.
We choose the objective function to be the fidelity 
$F_c={\rm Tr}\left[\rho_{c}(t)\rho_{W^{-}}\right]$, where
$\rho_{W^{-}}=|W^{-}\rangle\langle W^{-}|$, and choose the feedback control
Hamiltonian to be
$H_{\rm fb}=f_{1}\sigma^{x}_{1}+f_{2}\sigma^{x}_{2}+f_{3}\sigma^{x}_{3}$,
i.e., only single qubit rotations.  
The strategy  \cite{Ahn2002} to determine the feedback strengths
$f_{j}$ at each point in time is chosen optimally by maximizing
the fidelity $F_c={\rm Tr}\left[\rho_{c}(t)\rho_{W^{-}}\right]$.
By considering the dynamics of the feedback Hamiltonian only,
$\dot{\rho}_c(t)=-i[H_{\rm fb}, \rho_c(t)]$, 
the time evolution of the fidelity is
\begin{eqnarray}
\hspace{-0.3cm}\frac{dF_{c}(t)}{dt} &=&{\rm Tr}[\dot{\rho}_c(t)\rho_{W^{-}}]
=\left\langle-\imath\left[\rho_{W^{-}},
    H_{\rm fb}\right]\right\rangle_{c}\nonumber \\
&=&f_{1}\left\langle-\imath\left[\rho_{W^{-}}, \sigma_{1}^{x}\right]\right\rangle_{c}+f_{2}\left\langle-\imath\left[\rho_{W^{-}}, \sigma_{2}^{x}\right]\right\rangle_{c}+f_{3}\left\langle-\imath\left[\rho_{W^{-}}, \sigma_{3}^{x}\right]\right\rangle_{c}.\label{eq:Ft}
\end{eqnarray}
To maximize the fidelity, i.e., to make $[{dF_{c}(t)}/{dt}]$ positive
and maximal, 
the optimal feedback coefficients to kick the qubit system back to the
desired state $\rho_{W^{-}}=|W^{-}\rangle\langle W^{-}|$ are determined by 
$f_{j} =  f{\rm sgn}\left(\langle
  -\imath\left[\rho_{W^{-}},\sigma_{j}^{x}\right]\rangle_{c}\right)$,
where sgn($y$) denotes the sign function that extracts the sign of a real
number $y$, and $f$ is the maximum feedback strength that can be applied. 
This is a bang-bang feedback control scheme, meaning that the feedback
strengths are always at the maximum or minimum values \cite{Ahn2002}.

\subsection{Entanglement Creation and Stabilization }
\label{sec:fidelity}

\begin{figure}
\centering{}\includegraphics[width=0.45\columnwidth]{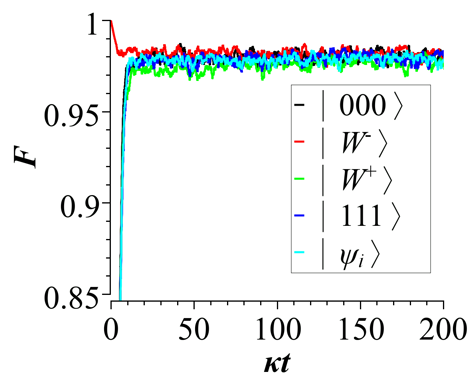} \caption{\label{fig:initial} (Color online)
  Time evolutions of the average fidelity $F$ of
  $\left|W^{-}\right\rangle$ state (over 1000 
  realizations or trajectories) as a function
  of time for different initial states of $\left|000\right\rangle$,
$\left|W^{-}\right\rangle$, $\left|W^{+}\right\rangle$,
$\left|000\right\rangle$ and $\left|\psi_i\right\rangle$. 
The parameters used are $f=2\kappa$, $\epsilon=2\kappa$, $\chi=-0.11\kappa$,
$g=10\kappa$, $\eta=1$, and $\gamma_j=\gamma=4\times 10^{-3}\kappa$.
}
\end{figure}

Considering qubits' decay rates $\gamma_j=\gamma=\kappa/250$ as in Ref. \cite{Lalumi`ere2010},
we demonstrate in Fig. \ref{fig:initial} that the entangled
$\left|W^{-}\right\rangle$ state can be generated and stabilized with
high average fidelity $F\approx0.98$ for various initial qubits' states by
our feedback control strategy with a moderate feedback strength of $f=2\kappa$.
The average fidelity $F$ is obtained by averaging $F_c$ over $1000$
realizations or trajectories. 
The various initial states are $|000\rangle$, $|W^{-}\rangle$,
$|W^{+}\rangle$, $|111\rangle$, eigenstates of the joint
measurement operator $c_{0}$, and the separable state
$|\psi_i\rangle$ of Eq.~(\ref{eq:JMOp2}). They all reach the average
 fidelity of about $0.98$ in a time
 scale of about a few $1/\kappa$, in about the same time scale for the 
cavity field to evolve into distinguishable coherent states (without
feedback control) in Fig.~\ref{fig:1}. Other parameters used in
Fig.~\ref{fig:initial} are the coupling strength $g=10\kappa$, and the
dispersive coupling strength $\chi=-0.11\kappa$.

\begin{figure}
\centering{}
\subfigure[][]{
\label{fig:Frelax_gamma}
\includegraphics[width=0.45\linewidth]{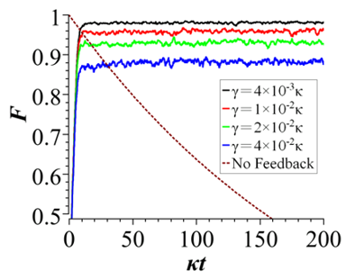}
}
\subfigure[][]{
\label{fig:Frelax_single}
\includegraphics[width=0.45\linewidth]{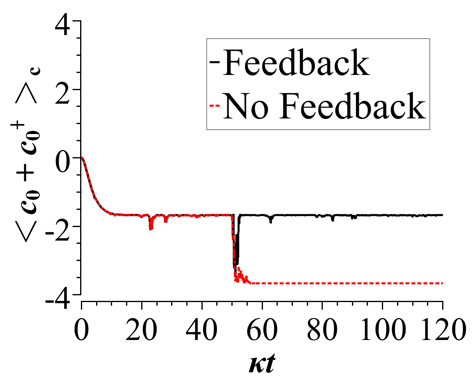}
}
\caption{\label{fig:Frelax} (Color online) (a) Time evolutions of the average
  fidelity $F$ of the 
  $\left|W^{-}\right\rangle$ state (over 1000
  trajectories) generated from the ground state
  $\left|000\right\rangle$ for different qubits' decay rates of
  $\gamma_j=\gamma$ being $4\times 10^{-3}\kappa$, $1\times
  10^{-2}\kappa$, $2\times 10^{-2}\kappa$, and $4\times 10^{-2}\kappa$
  (solid lines from top to bottom). 
 The brown-dashed line is the ensemble average 
  fidelity without the application of feedback control for an initial qubits' state being in $\left|W^{-}\right\rangle$. 
(b) A typical single trajectory of the estimated current $\langle
 c_{0}+c^{\dagger}_{0}\rangle_c$  for the qubits' decay rates of
  $\gamma_j=\gamma=4\times 10^{-3}\kappa$.
 The qubits' system is stabilized in the $|W^{-}\rangle$ state with $\langle
 c_{0}+c^{\dagger}_{0}\rangle_c=-1.68$ during the 
continuous feedback
 control process (in solid line), while it makes a sudden jump to the
 ground state $\left|000\right\rangle$ with $\langle
 c_{0}+c^{\dagger}_{0}\rangle_c=-3.68$ without the
 application of feedback
 control (in dashed line.
Other parameters used are the same as those in Fig.~\ref{fig:initial}. 
}
\end{figure}

Figure \ref{fig:Frelax}(a) shows the time evolutions
of the average fidelity of the $\left|W^{-}\right\rangle$ state for different
qubit decay rates. The brown-dashed line is the ensemble average 
(unconditional) result of the qubit system evolving from the
$\left|W^{-}\right\rangle$ state without feedback control for the
qubit decay rates of $\gamma_j=\gamma=4\times 10^{-3}\kappa$.
The fidelity without feedback control deteriorates about linearly with
time. This can be understood from a typical measured current 
record in a single realization of the experiment
as shown in Fig. \ref{fig:Frelax}(b).
The qubits initially in the $\left|W^{-}\right\rangle$ state
corresponding to the result of $\langle
c_{0}+c^\dagger_{0}\rangle_c= -1.68$ for the parameters chosen here 
 have probability $\gamma dt$ to make a sudden jump into the ground state
$\left|000\right\rangle$ corresponding to $\langle
c_{0}+c^\dagger_{0}\rangle_c= -3.68$ in the time interval $\left[t,
  t+dt\right]$.
If there is no feedback control, the qubit 
after the sudden jump
 will then stay in the ground
state as indicated by the red-dashed curve in Fig.~\ref{fig:Frelax}(b).
In contrast, even when the initial qubit state is the ground state
$\left|000\right\rangle$, the qubit with feedback control will be
driven  to the
$\left|W^{-}\right\rangle$ state and be stabilized for a sufficient
long time. This is also shown in Fig. \ref{fig:Frelax}(a) in which 
the ensemble averaged fidelities with application of 
feedback control (even when the decay
 rate of the qubits is $\gamma_j=\gamma=\kappa/25=4\times
 10^{-2}\kappa$) are stabilized with 
values above that of the brown-dashed line (with
$\gamma_j=\gamma=\kappa/25$) when time $\kappa t > 50$.

\begin{figure}
\centering
\subfigure[][]{
\label{fig:fine_chi}
\includegraphics[width=0.45\columnwidth]{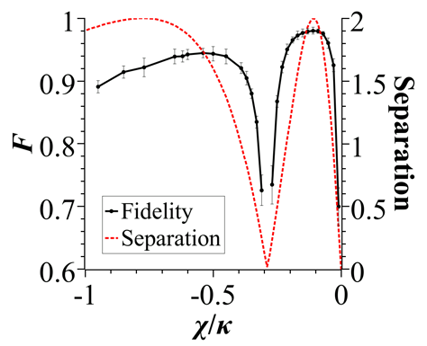}
}
\subfigure[][]{
\label{fig:fine_local}
\includegraphics[width=0.45\columnwidth]{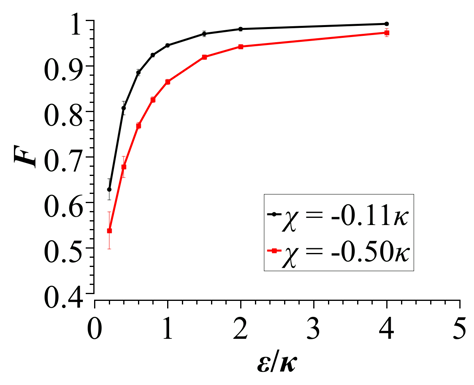}
}

\subfigure[][]{
\label{fig:fine_control}
\includegraphics[width=0.45\columnwidth]{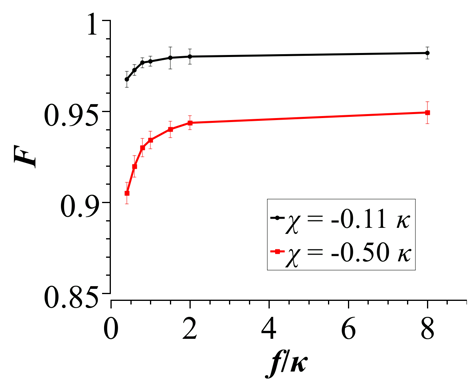}
}
\subfigure[][]{
\label{fig:fine_ineff}
\includegraphics[width=0.45\columnwidth]{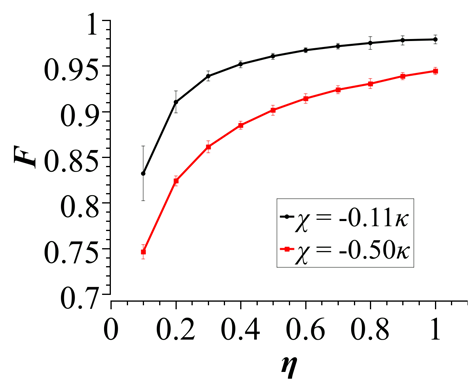}
}

\caption{\label{fig:fine} (Color online) Dependence of the average
    fidelity $F$ of the 
  $\left|W^{-}\right\rangle$ state at $\kappa t=350$ on
\subref{fig:fine_chi} the dispersive coupling strength $\chi$, 
\subref{fig:fine_local} the driving amplitude $\epsilon$, 
\subref{fig:fine_control} the feedback strength $f$ and 
\subref{fig:fine_ineff} the measurement efficiency $\eta$. 
The decay rate of the qubits is fixed at 
$\gamma_j=\gamma=4\times 10^{-3}\kappa$ and the initial qubit state is the ground state $\left|000\right\rangle$.   
The red-dashed curve in \subref{fig:fine_chi}
is the separation between the measurement
outcomes of the ground state and the $\left|W^{-}\right\rangle$ state
(or the single-qubit excited state) with its vertical axis label
shown on the right.}

\end{figure}

We discuss in the following the dependence of the average fidelity 
on the dispersive coupling strength
$\chi$, the probe beam amplitude $\epsilon$, the feedback strength
$f$, and the measurement efficiency $\eta$. 
The black-circle solid line in Fig.~\ref{fig:fine}\subref{fig:fine_chi} is
the average fidelity $F$ versus the dispersive coupling strength $\chi$ for the
probe field $\epsilon=2\kappa$, the feedback strength $f=2\kappa$, and
the decay rate of the qubits $\gamma_j=\gamma=4\times10^{-3}\kappa$. The
dependence of the fidelity on $\chi$ is similar to the red-dashed
curve which
represents the separation between the measurement output signal $\langle
c_{0}+c^\dagger_{0}\rangle_c$ that corresponds to the qubits' state being in
$\left|W^{-}\right\rangle$ and the output signal that corresponds to
the qubits' state of 
$\left|000\right\rangle$. This is because larger separation means
better state distinguishibility and thus helps the conditional qubits'
state estimation in the quantum feedback control scheme. 
One can observe that $\left|W^{-}\right\rangle$ and 
$\left|000\right\rangle$ become indistinguishable from the measurement
current around the point $\chi\approx-0.29\kappa$, and thus the fidelity
drops sharply around there as well. 
One can also notice from
  Fig.~\ref{fig:fine}\subref{fig:fine_chi} that 
the separation reaches maximum values at $\chi=-0.11\kappa$ and
$\chi=-0.77\kappa$, 
but the fidelity is higher at $\chi=-0.11\kappa$.
This is because the collective damping rate $\kappa\lambda^{2}=\kappa\chi^2/g^2$
of the Purcell effect
of the third term in Eq.~(\ref{eq:MS}) increases with the values of
$\chi$. 
For example, for $g=10\kappa$, the Purcell collective damping rate
of $5.93\times 10^{-3}\kappa$ at $\chi=-0.77\kappa$ 
is larger than the individual
qubit decay rate, set to be $\gamma_j=\gamma=4\times10^{-3}\kappa$
here, while the Purcell collective damping rate
of $1.21\times 10^{-4}\kappa$ 
at $\chi=-0.11\kappa$ is much smaller than $\gamma$ and thus does not
play an important role.
As a result, the average fidelity is lower for the case of 
a higher $\chi$ value when the corresponding separation of the measurement 
outcomes is the same.
Another observation from Fig.~\ref{fig:fine}\subref{fig:fine_chi} is
that the average fidelity does not change with 
the dispersive coupling strength $\chi$ as sharply as the
separation of the measurement outcomes does. 
When the separation of the measurement outcomes above a certain value (about
$1.6$ for the parameters chosen here), the average fidelity does not
vary much [see the behaviors of 
the separation of the measurement outcomes 
and the average fidelity around $\chi=-0.11\kappa$, where the Purcell
effect is not significant as compared to the individual qubits' decay]. 
This may also explain why the average fidelity
at $\chi=-0.5\kappa$ is larger than that at $\chi=-0.77\kappa$. 
The Purcell collective damping rate of $2.50\times 10^{-3}\kappa$ at
$\chi=-0.5\kappa$, which is smaller than the individual
qubit decay rate $\gamma_j=\gamma=4\times10^{-3}\kappa$, is smaller
than that of $5.93\times 10^{-3}\kappa$ at $\chi=-0.77\kappa$. 
Although the separation of the measurement outcomes at $\chi=-0.5\kappa$
is also smaller there, its value
 is larger than $1.6$. 
As a result, the average fidelity
at $\chi=-0.5\kappa$ is larger than that at $\chi=-0.77\kappa$.
We perform most of our simulations choosing $\chi=-0.11\kappa$ and/or
$\chi=-0.5\kappa$. 


The dependence of average fidelity on the measurement drive amplitude
$\epsilon$ is shown in
Fig.~\ref{fig:fine}\subref{fig:fine_local}.
Since the information gain rates [c.f.~Eqs.~(\ref{eq:rate0}) and
(\ref{eq:rate1})] is proportional to $\sqrt{\Gamma_{\rm m}}\propto \epsilon$, 
the bigger the value $\epsilon$ is, the larger
the separation between measurement outcomes is and the quicker the
conditional state collapse to one of the joint measurement operator eigenstates
is. It is thus expected the average fidelity will also become better
as $\epsilon$ increases as shown in Fig.~\ref{fig:fine}\subref{fig:fine_local}. 
One may be temped to think that the arbitrarily quick readout or
arbitrarily high fidelity can be achieved by simply increasing
$\epsilon$. 
But it was pointed out \cite{Blais2004,Gambetta2006} 
that the lowest-order dispersive approximation
of Hamiltonian Eq.~(\ref{eq:effHamiltonian}) become accurate when the
average photon number in the cavity is much smaller than the critical
photon number of $n_{\rm crit}=\Delta^2/4g^2$. The number of photon is
proportional to $\epsilon^2$. This puts a limit on how large the
external drive $\epsilon$ could be for  Eq.~(\ref{eq:effHamiltonian})
to hold valid. 
In addition, note that the time-dependent second term $\sum_{j}\epsilon\lambda(\sigma_{j}^{+}e^{i\Delta t}+\sigma_{j}^{-}e^{-i\Delta t})$
in the first commutator of
Eq.~(\ref{eq:ME}) also increases with $\epsilon$. This term in the
Hamiltonian, in addition to qubit decay channel, will cause
the qubits to flip or change their state during the process when the continuous measurement
tries to localize the qubits 
to one of the joint measurement operator eigenstates.
However, the value of $\lambda=\chi/g=0.11/10=0.011$ we choose is small and 
for typical value of $\epsilon$, the coefficient $\epsilon\lambda$ of
this term is much smaller than the
frequency $\Delta=\Omega-\omega_d$ (as we have set
$\omega_d=\omega_r$) of the oscillating factors.
Thus the
effect of this term to mix different measurement eigenstates is small.
We choose
 $\epsilon=2\kappa$ for most of the simulations presented in this
 paper although increasing $\epsilon$ further will improve the
 fidelity a little bit.

Figure \ref{fig:fine}\subref{fig:fine_control} shows that  
increasing
the feedback control strength improves the average fidelity in
general.
Suppose a measurement outcome indicating deviation from the desired
$\left|W^{-}\right\rangle$ state happens, the applied feedback control has
to overcome the effect of the localization due to the continuous
measurement in order to move the qubits back to the target
$\left|W^{-}\right\rangle$ state. When the feedback control
strength is smaller, the procedure to produce and stabilize the
$\left|W^{-}\right\rangle$ state takes a
longer time with a lower fidelity.  
The qubits' decay rates in Fig.~\ref{fig:fine}\subref{fig:fine_control}
are chosen to be $\gamma_j=\gamma=4\times 10^{-3}\kappa$.
When the feedback control
strength is $f=2\kappa$, the fidelity reaches the value of $0.98$.
Further increase of the feedback control
strength, i.e., $f>2\kappa$, does not improve appreciably the fidelity. 
This indicates that if deviation occurs, the correction of the feedback control
at $f=2\kappa$ is fast
enough to kick the qubits back to the $\left|W^{-}\right\rangle$ state.
We thus choose $f=2\kappa$ for most of our numerical simulations.

In practice, there exists inefficiency in the measurements which arises
when the
detectors sometimes miss detection or the measurement microwave
photons does not go to the
detectors due to lost. 
However, high measurement efficiency
is not very essential for our feedback control scheme. 
Although the fidelity decreases as the value of the measurement efficiency
$\eta$ decreases as shown in Fig. \ref{fig:fine}\subref{fig:fine_ineff}, the fidelity is still above $0.9$ for $\eta$ as low
as $0.2$ for the case of $\chi=-0.11\kappa$ and 
qubits' decay rates $\gamma_j=\gamma=4\times 10^{-3}\kappa$.
The value of $\eta<1$ implies appearance of an additional
non-unraveling dephasing term in the quantum trajectory (stochastic
master)
equation \cite{Tornberg2010}. However, this term causes dephasing only among   
$\left|111\right\rangle$,
$\left|W^{+}\right\rangle$, $\left|W^{-}\right\rangle$ and
$\left|000\right\rangle$ for the initial qubits' states chosen in our
simulations.  
As a result, it affects only the detailed dynamics of the qubits 
but does not destroy or prevent the controlled evolution toward the
target entangled state
$\left|W^{-}\right\rangle$. Therefore, our feedback control scheme to generate 
and stabilize the $\left|W^{-}\right\rangle$ state does not
require high measurement efficiency and thus can be
implemented experimentally with high fidelity with measurement
efficiency available in current circuit QED experiments
  that use parametric amplification before the homodyne detection with
  an IQ mixer
  \cite{Castellanos-Beltran08,Vijay2011,Vijay2012,Mallet11}. 
 For example, Refs.~\onlinecite{Vijay2012,Mallet11} have achieved an effective quantum efficiency of $\eta=0.4$.

\begin{figure}
\centering
\includegraphics[width=0.45\columnwidth]{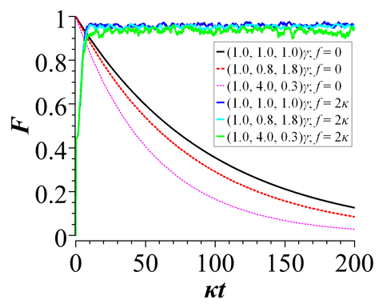}
\caption{\label{fig:9} 
(Color online) Time evolutions of the average fidelity $F$ of the 
  $\left|W^{-}\right\rangle$ state with ($f=2\kappa$) and without ($f=0$)
  quantum feedback control for three sets of different individual
  qubits' decay rates. 
The initial state for the case with feedback control
  ($f=2\kappa$) is the ground state $|000\rangle$ while it is 
the  $\left|W^{-}\right\rangle$ state for the case without feedback control ($f=0$).
Other parameters used are $\epsilon=2\kappa$, $\chi=-0.11\kappa$,
$g=10\kappa$, $\eta=1$, and $\gamma=10^{-2}\kappa$.}

\end{figure}

Our feedback control scheme is robust even when the decay rates of the
qubits are different. This is shown in Fig.~\ref{fig:9} where the time
evolutions of the 
average fidelity of the 
  $\left|W^{-}\right\rangle$ state with and without feedback control for three
sets of qubits' decay rates are shown. 
The average fidelity is determined roughly
by the average decay rate in each set as the behavior of the
fidelity in each set is
similar to that when the decay rates of the three qubits were
equal to the average decay rate.   
The average fidelities of the 
  $\left|W^{-}\right\rangle$ state generated initially from the ground state
  $|000\rangle$ with feedback control strength 
  $f=2\kappa$  outperform those
 evolving from an initial
  $\left|W^{-}\right\rangle$ state without feedback control ($f=0$)
after time $\kappa t\approx 15$.

\section{Comparison with adiabatic elimination method}
\label{sec:comparison} 

Another commonly used procedure to eliminate the cavity field is the
adiabatic elimination method. 
Both the adiabatic method and the polaron-type
transformation method assume that $\kappa\gg \gamma_i$, but    
 the adiabatic method employs an additional condition, i.e., to assume
that the damping of the cavity is much larger than
the dispersive coupling strength, i.e., $\kappa\gg \chi_i$. 
In the limit of $\kappa\gg\chi$, 
the term $\chi_{x}$ in Eq.~(\ref{eq:Coherent}) is ignored 
and the TLR cavity field reaches its steady coherent state rapidly
with an amplitude equal to $\alpha=-2\imath\epsilon/\kappa$.
As a consequence, the coherent state amplitude is assumed to be the same
for all the qubits' basis states.
This is in contrast to the case in
the polaron-type transformation method where the time-dependent coherent state
amplitudes $\alpha_x$  shown in
Eq.~(\ref{eq:Coherent}) depend on $\chi_x$ and thus on the qubits basis
states $|x\rangle$.
The steady-state information gain rates in the limit of $\kappa\gg\chi$ from Eqs.~(\ref{eq:rate0}) and
(\ref{eq:rate1}) become $\sqrt{\kappa\Gamma_0}\to\sqrt{\Gamma_{\rm
    m}}$ and $\Gamma_1\to 0$ (note also that $\Gamma_2\to 0$).
As a result, the measurement operator from Eq.~(\ref{eq:JMOp2})
  becomes $\sqrt{\Gamma_{e}/\kappa}\sum_{j}\sigma_{j}^{z}$.
The effective conditional (stochastic) master equation 
(\ref{eq:SME}) in the case of adiabatic
elimination also reduces to 
\begin{eqnarray}
\frac{d\rho_{c}^{\text{e}}\left(t\right)}{dt} & = & 
-\imath\left[\sum_{j}[({\chi}/{2})+\chi\left|\alpha\right|^{2}]\sigma_{j}^{z}
+\sum_{j}\epsilon\lambda(\sigma_{j}^{+}e^{i\Delta
  t}+\sigma_{j}^{-}e^{-i\Delta t})
+\sum_{j>i}\chi\left(\sigma_{i}^{-}\sigma_{j}^{+}+\sigma_{i}^{+}\sigma_{j}^{-}\right),
\rho_{c}^{e}\left(t\right)\right]\nonumber\\
&&+\sum_{j}\gamma_{j}\mathcal{D}\left[\sigma_{j}^{-}\right]\rho_{c}^{e}\left(t\right)+\kappa\mathcal{D}\left[\sum_{j}\lambda\sigma_{j}^{-}\right]\rho^{e}_{c}\left(t\right)\nonumber\\
&&+\frac{\Gamma_{\rm e}}{2}\mathcal{D}\left[\sum_{j}\sigma_{j}^{z}\right]\rho^{e}_{c}\left(t\right)
+\frac{\sqrt{\eta \Gamma_{\rm m}}}{2}\mathcal{H}\left[\sum_{j}\sigma_{j}^{z}\right]\rho^{e}_{c}\left(t\right)\xi\left(t\right).\label{eq:aSME}
\end{eqnarray}
Here, the second term in the first commutator term and the fourth
term of Eq.~(\ref{eq:aSME}) are reduced respectively from Eqs.~(\ref{eq:acStark}) and
(\ref{eq:dephasing}) of Eq.~(\ref{eq:MS}). 
Note again that the measurement rate here is twice of the
  decoherencee rate, $\Gamma_{\rm m}=2\Gamma_{\rm e}=64\epsilon^{2}\chi^{2}/\kappa^{3}$.
One can clearly see that the adiabatic elimination procedure is a
special case of polaron-type transformation in the limit of 
$\kappa\gg \chi$.

The measurement outcomes of the average homodyne currents 
obtained by categorizing and averaging $1000$ realizations that
yield roughly the same steady outcome values for the case of
adiabatic elimination are 
plotted in Fig.~\ref{fig:1} to compare with the case of polaron-type
transformation with the same parameters. 
The qubits are initially in the separable state
$\left|\psi_{i}\right\rangle$ of Eq.~(\ref{eq:FactorState}) with the qubits'
decay rates set to zero, i.e., $\gamma_j=0$, and 
the cavity state evolves from an initial vacuum state. 
One can see that the measurement outcomes in dashed lines for the adiabatic
elimination case approach to their corresponding
steady values more 
quickly. Moreover, the four measurement 
outcomes in the adiabatic elimination limit become $3\sqrt{\Gamma_{\rm m}/\kappa}$,
$\sqrt{\Gamma_{\rm m}/\kappa}$, $-\sqrt{\Gamma_{\rm m}/\kappa}$,
and $-3\sqrt{\Gamma_{\rm m}/\kappa}$; as a result, the steady value 
corresponding 
to $\left|111\right\rangle$ ( $\left|000\right\rangle$) is
overestimated, i.e., becomes larger (smaller). Thus neglecting the
contribution of $\Gamma_1\to 0$ in the case of adiabatic elimination for
the parameter of $(\chi/\kappa)=-0.11$ used in Fig.~\ref{fig:1} is not
really valid. 

For the adoptive feedback control by state estimation
  method, it is important to use the correct conditional stochastic
  master equation to estimate the system state conditioned on the
  measured current. Otherwise, wrong state estimation information will give rise
  to bad feedback control result. We have tested numerically that when $|\chi/\kappa|\le 0.01$ 
both the adiabatic elimination method and the polaron-type
transformation method give the same result for the typical parameters
chosen in our simulation in the absent of feedback control.
However, when $|\chi/\kappa|> 0.02$, discrepancy in conditional qubits'
trajectories starts to emerge. 
However, 
Fig.~\ref{fig:fine}\subref{fig:fine_chi} indicates that in the
presence of feedback control the average fidelity is below $0.8$ for
$\gamma_j=\gamma=\kappa/250$ with this value of $|\chi/\kappa|< 0.02$. 
For the value of $(\chi/\kappa)= -0.11$, the fidelity to
stabilize the $|W^{-}\rangle$ can be maintained at $0.98$ using the
feedback control master equation (\ref{eq:feedbackSME})
obtained by the polaron-type
transformation method.
In other words, in the parameter regime where the adiabatic elimination does not apply, the conditional stochastic master equation (\ref{eq:aSME}) can not be used; otherwise the wrong information about the system state will lead to low-fidelity feedback control results.
If one would use the conditional master equation
(\ref{eq:aSME}) obtained by the adiabatic elimination method
to perform the feedback control scheme calculation by adding a
feedback Hamiltonian commutator term  for the case of 
$(\chi/\kappa)= 0.11$, high average fidelity of $0.95$ could be achieved.
But this is not correct as the conditional master equation
(\ref{eq:aSME}) is not really valid when $(\chi/\kappa)=-0.11$. 
In fact, if we nevertheless use the conditional master equation
(\ref{eq:aSME}) obtained by the adiabatic elimination method for the state estimation to determine the sign of the
feedback strength and then use the 
polaron-type feedback control
master equation (\ref{eq:feedbackSME}) (mimicking the real experimental
situation) to  
evolve and calculate the fidelity, 
the average fidelity is found to be below $0.5$. 
This is because the signs of the feedback control strength 
estimated by Eq.~(\ref{eq:aSME}) obtained by the adiabatic elimination
method for the value of $(\chi/\kappa)= -0.11$ are often wrong.


\section{Conclusion}
\label{sec:conclusion}

In conclusion, we have presented a simple and promising quantum
feedback control scheme for deterministic generation and stabilization
of a three-qubit $|W^-\rangle$ state in a superconducting circuit QED setup,
taking into account the realistic conditions of decoherence and decay.
Our scheme is based on continuous joint Zeno measurements of
multiple qubits in a dispersive regime and the application of 
multi-qubit adaptive feedback control. 
The dispersive measurement not only enables qubit state estimation 
for further information processing but also allows, together with the
feedback control, for the generation and stabilization of the target
entangled $|W^-\rangle$ state starting from separable input states or
from the ground states of the qubits. 
The feedback control Hamiltonian can be
realized by applying, besides the measurement drive, an additional
control microwave drive with a frequency in resonance with the qubits'
transition frequency. 
We have employed the polaron-type transformation
method to eliminate
the cavity field to obtain an effective stochastic master equation 
for the qubits' degrees of freedom alone, and
simulated the dynamics of the proposed quantum feedback
control scheme using the quantum trajectory approach. 
It is demonstrated
that in the presence of moderate environmental decoherence,
the average entangled state fidelity higher than $0.9$ can be achieved and
maintained for a considerable long time (much longer than the qubits'
decoherence time) with our scheme. 
In our discussion, we have assumed to have identical
  qubits with transition frequencies
  $\Omega_{1}=\Omega_{2}=\Omega_{3}=\Omega$ and couplings
  $g_{1}=g_{2}=g_{3}=g$. Although the couplings $g_j$  of the qubits
  to the cavity field may, in a realistic experiment, not be the same,
  one is able to tune, due to their great tunability, the qubit
  transition frequencies to be pretty much the same by external
  voltages or magnetic fields. In other words, experimentally the
  detuning $\Delta=\Omega-\omega_r$ for all the qubits can be tuned to
  be equal while the dispersive coupling strengths
  $\chi_j=g_j^2/\Delta$ are left slightly different. We have tested
  numerically that a mismatch smaller than 
  $10^{-3}g=10^{-2}\kappa$ (we set $g=10\kappa$) in the coupling strengths $g_j$  changes
  insignificantly the fidelity to achieve the desired
  $|W^-\rangle$ state. However, a mismatch of $5\times10^{-2}\kappa$
  ($10^{-1}\kappa$) in $g_j$ results in a fidelity change, say for
  $\gamma=4\times10^{-3}\kappa$ case, from 0.98 to 0.93 (0.84). Taking
  the values of $\kappa$ to be about $5$ MHz yields a tolerant
  mismatch in $g_j$, which will not affect the desired fidelity, to be about $10^{-2}\kappa=0.05$ MHz. 
The required values for the physical parameters are 
achievable in current experiments.
Our method can also be extended straightforwardly to generate and stabilize an
  $N$-qubit (with $N>3$) W-type state with one
 excitation shared across $N$ qubits in superposition.

We have also compared the polaron-type transformation
method with the adiabatic elimination method to eliminate
the cavity field. It is shown that the adiabatic elimination procedure is a
special case of polaron-type transformation in the limit of 
$\kappa\gg \chi$. Our feedback control scheme is also shown to be
robust against measurement inefficiency and individual qubit decay
rate differences.  
Recently, quantum feedback experiments stabilizing Fock states of
light in a cavity by using sensitive atoms, crossing the field one at
a time as quantum nondemolition probes of its photon number have been
reported \cite{Haroche}.
An experiment of stabilizing Rabi
oscillation of a superconducting qubit in a cavity 
using quantum feedback control via homodyne measurements
has also been demonstrated \cite{Vijay2012}. 
Although the measurement efficiency  was estimated
to be about $0.4$ in the quantum feedback control experiment of a
superconducting qubit, the Rabi oscillations was shown to 
persist indefinitely. Our feedback control scheme to generate and
stabilize entangled state can still achieve high fidelity even when
measurement efficiency is as low as $0.4$. 
Furthermore, processing data  in real time using fast field-programmable gate
array (FPGA) electronics in circuit QED setup 
has been demonstrated \cite{Bozyigit2011}, and this will facilitate the performance of quantum state
estimation in real time in our scheme. 
Thus our quantum feedback scheme has great potential to be
realized experimentally in the near future.

\begin{acknowledgments}
HSG acknowledges support from the
National Science Council in Taiwan under Grant
No.~100-2112-M-002-003-MY3, 
from the National Taiwan University under Grants
No.~102R891400, No.~102R891402 and No.~102R3253, and
from the
focus group program of the National Center for Theoretical
Sciences, Taiwan.
\end{acknowledgments}

\end{document}